\documentclass[final,5p,times,twocolumn]{elsarticle}
\usepackage{lineno}
\usepackage{graphicx}
\biboptions{numbers,sort&compress} 
\usepackage[integrals]{wasysym} 
\modulolinenumbers[5]
\usepackage{natbib}

\usepackage{etoolbox}
\AtBeginEnvironment{thebibliography}{%
  \setlength{\parskip}{0pt}%
  \setlength{\itemsep}{0pt}%
}

\usepackage{booktabs,siunitx,threeparttable}
\usepackage{amssymb, hyperref}
\usepackage{amsmath}
\usepackage{lineno}
\usepackage{booktabs}
\usepackage{siunitx}
\usepackage{threeparttable}
\usepackage{makecell}

\begin{document}

\begin{frontmatter}

\title{Collision and static widths of self-broadened line shapes in optically 
saturated high-density atomic vapor}

\author[JIHT]{Vladimir Sautenkov}
\author[JIHT,HSE]{Sergey Saakyan\corref{cor}}
\ead{saakyan@ihed.ras.ru}
\author[JIHT]{Andrei Bobrov}
\author[JIHT]{Eugenia Vilshanskaya}
\author[JIHT]{Boris B. Zelener}

\address[JIHT]{Joint Institute for High Temperatures, Russian Academy of 
Sciences (JIHT RAS), Izhorskaya St. 13 Bld. 2, Moscow 125412, Russia}
\address[HSE]{National Research University Higher School of Economics (NRU HSE),
Myasnitskaya Ulitsa 20, Moscow 101000, Russia}
\cortext[cor]{Corresponding author}

\begin{abstract}
We study the frequency derivatives of selective reflection from high-density rubidium 
vapor using a hole-burning technique. 
Saturation dips are observed inside the self-broadened line shapes. 
The line self-broadening is a combination of static width and collision width. 
By analyzing saturation dips, we can separate power broadening and collision width. 
Our experimental results support the theory of inhomogeneous dipole-dipole induced 
broadening of transitions in a dense atomic gas, published 
by J.~A. Leegwater and S. Mukamel [Phys. Rev. A 49 (1994) 146].

\end{abstract}

\begin{keyword}
    rubidium atoms\sep dense atomic gas\sep self-broadening\sep many-body effects
\end{keyword}

\end{frontmatter}


\section{Introduction}
The standard theory of line self-broadening in atomic gas is based on a concept
of two particle dipole-dipole interactions~\cite{Lewis1980}. 
The dipole induced broadening $\Gamma_0$ depends linearly on the atomic density $N$ 
\begin{equation}\label{eq1}
  \Gamma_0 = K N,
\end{equation}
where the factor $K/2\pi \approx 10^{-7}$~Hz\,cm$^3$~\cite{Lewis1980}. 
In a high-density atomic gas, where the line self-broadening of the resonance 
transitions dominates over the Doppler width, the thermal motion of atoms can be 
neglected. 
The two-particle approximation must be modified for a correct description of 
high-density atomic gas media.

The theory of line self-broadening has been extended within the framework of disordered 
excitons~\cite{MukamelPRA1994}. 
In the proposed model, the interacting particles have infinite mass. 
It is shown that, in the analysis of the line self-broadening the many particle effects have 
to be considered. 
In the literature spectral profiles of resonance transitions are usually classified as either 
homogeneous or inhomogeneous. 
A spectral line shape is considered to be homogeneously broadened when the main broadening 
mechanism is induced by the binary collisions of the moving atoms (collision broadening), 
and inhomogeneous if the spectral width is induced by a distribution of transition 
frequencies (static broadening). 
The inhomogeneous self-broadening due to many-particle interaction is described by the 
developed model of frequency shifted excitons with long-range interactions. 
The concept of the Lorentz local field in a disordered medium is also applied.
In linear optics, it is impossible to distinguish between the two broadening mechanisms
in self-broadened line shapes in a dense atomic gas. 
In the developed approach, both broadening mechanisms produce Lorentzian-like
spectral profiles. 
The calculated ratio of static and collision widths is independent of the atomic 
density.

The homogeneous or inhomogeneous character of the spectral profile of an atomic transition
can be tested by nonlinear laser spectroscopy methods such as photon echoes or 
hole burning~\cite{MukamelJCP1991}. 
Measurements of self-broadening in high-density potassium vapor are 
satisfactorily described by the standard theory~\cite{BoydPRL1991}.

Application of a nonlinear optical technique to the investigation of line-shapes in 
high-density atomic vapor is reported in~\cite{VASPRL1996}. 
In the experiment with rubidium atomic vapor, the probe and pump tunable diode lasers 
are used. 
The probe laser frequency is tuned over the $5S_{1/2}-5P_{3/2}$ transition ($D_2$ line).
The pump laser frequency is fixed in the far wing of the atomic transition, where the 
absorption length is much longer than the wavelength. 
The off-resonant optical excitation of rubidium atoms is incoherent due to reabsorption
and re-emission of photons, together with nonradiative excitation transfer 
processes~\cite{HolsteinPRA1951, Biberman1947, VASJQSRT2020}. 
By analyzing the experimental and calculated reflection spectra, a linear dependence of the line 
self-broadening $\Gamma$ on the ground state population $N_g$ is obtained~\cite{VASPRL1996}. 
The observed excitation dependence is explained by the static mechanism of the dipole–dipole 
interactions in the high-density atomic vapor~\cite{MukamelPRA1994}. 
The results of subsequent publications, discussed in a review paper~\cite{VASreviewLPL2011}, 
confirm the linear relation between $\Gamma$ and $N_g$ in incoherently excited potassium 
and rubidium vapors.

The observation of resonant saturation of selective reflection from 
high-density rubidium vapor is discussed in~\cite{VASNSRJQSRT2024}. 
Using the pump-probe technique, the spectroscopic measurements are performed for the 
$D_2$-line at different atomic densities in the range $(1.2-3.6)\times10^{17}$~cm$^{-3}$.
The spectral dependence of the selective reflection coefficient for the weak probe beam is 
recorded at several pump beam intensities from $0$ to $8.8$~kW\,cm$^{-2}$. 
At high pump intensities, weak variations of the reflectivity appear near the 
pump beam frequency. 
These spectral features are explained by power broadening effects.

In the following work~\cite{VASJETPLett2025}, we record the derivative selective reflection 
coefficient $\mathrm{d}R/\mathrm{d}\nu$ at the maximum density $3.6\times10^{17}$~cm$^{-3}$. 
The use of the derivative $\mathrm{d}R/\mathrm{d}\nu$ allows us to improve the spectral resolution
and the signal to noise ratio. 
At high pump intensities, the spectra are split into two symmetric resonances. 
The splitting intervals are equal to twice the Rabi frequency. 
The observed narrow structures are associated with 
"dressed atomic states"~\cite{Cohen-TannoudjiJPhysB1977} in high-density atomic 
vapor.

In~\cite{VASJQSRT2026}, an investigation of nonlinear selective reflection from 
rubidium atomic vapor is presented. 
In the experiments, the hole-burning technique with probe and pump lasers is applied. 
The derivative $\mathrm{d}R/\mathrm{d}\nu$ is recorded and analyzed. 
Changing the atomic number density from $2.5 \times 10^{17}$ to 
$3.6 \times 10^{17}$~cm$^{-3}$ induces a change in the character of line self-broadening 
from inhomogeneous to homogeneous. 
At the highest density, the strong pump field splits the observed spectra into two 
homogeneously broadened symmetric resonances. 
The appearance of the narrow symmetric resonances can be explained by the 
"dressed atomic states" approach~\cite{Cohen-TannoudjiJPhysB1977}. 
In the lower density range $(1.2-2.5)\times10^{17}$~cm$^{-3}$, the spectral profiles 
are inhomogeneously broadened~\cite{MukamelPRA1994}. 
The optically induced dips are observed in the spectral profiles of the derivatives. 
The width of each saturation resonance is a combination of the homogeneous collision width 
and line power broadening.

The goal of the present research is to distinguish the collision and static widths of
self-broadened transitions in high-density atomic vapor.

\section{Experimental results and discussion}
The resonant saturation of selective reflection from the window--atomic vapor interface 
in a high-temperature optical cell with a natural abundance of $^{85}$Rb and $^{87}$Rb
isotopes is studied at the $5S_{1/2}-5P_{3/2}$ transition ($\lambda = 780$~nm). 
Using formula~(\ref{eq1}) for self-broadening with the factor 
$K/2\pi = (1.1 \pm 0.17) \times 10^{-16}$~GHz\,cm$^3$ for the selected 
transition~\cite{AdamsJPhysB2011}, the line widths $\Gamma_0$ are evaluated for three selected 
atomic densities $N_i=(1.2, 1.7, 2.5) \times 10^{17}$~cm$^{-3}$ (see Table~\ref{tab}).

\begin{figure}[h]
    \centering
    \includegraphics{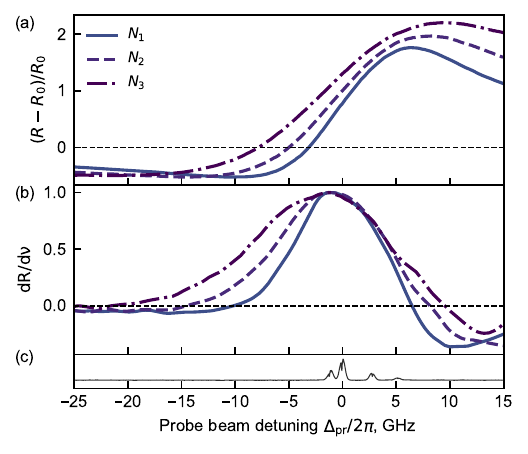}
    \caption{Selective reflection spectra recorded in the linear regime ($I = 0$).
    (a)~Normalized reflection coefficient $(R-R_0)/R_0$ for three atomic densities $N_i$.
    (b)~Frequency derivatives $\mathrm{d}R/\mathrm{d}\nu$ for the same three atomic densities.
    (c)~Reference saturated absorption spectrum measured in a rubidium vapor cell.}
    \label{fig1}
\end{figure}

\begin{figure*}[t!]
    \centering
    \includegraphics[width=\linewidth,height=0.36\textheight,keepaspectratio]{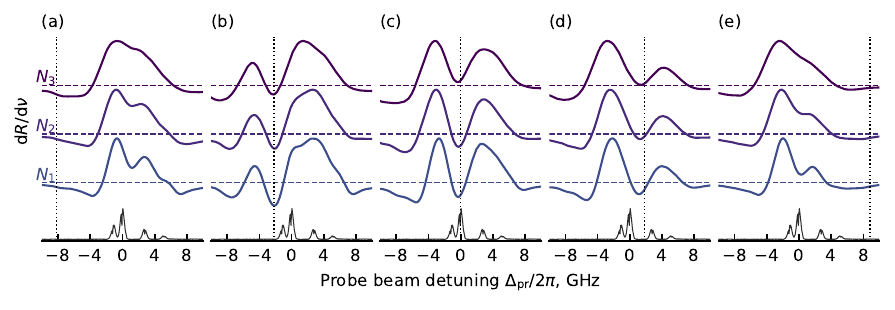}
    \caption{Derivatives $\mathrm{d}R/\mathrm{d}\nu$ of the selective reflection spectra
      measured at the maximum pump intensity, $I=8.8$~kW/cm$^2$, for different pump laser 
      detunings of (a)~$-8.2$~GHz, (b)~$-2.2$~GHz, (c)~$0$~GHz, (d)~$1.8$~GHz, and 
      (e)~$8.8$~GHz.
      The vertical dotted lines indicate the pump beam detunings relative to the
      $5S_{1/2}(F=3)-5P_{3/2}(F'=4)$ hyperfine transition of
      $^{85}\mathrm{Rb}$ in the reference cell. 
      Each row corresponds to a different number density $N_i$, indicated on the left, 
      increasing from the lowest density, $N_1=1.2\times10^{17}$~cm$^{-3}$, 
      to the highest density, $N_3=2.5\times10^{17}$~cm$^{-3}$. 
      The bottom row shows the reference saturated absorption spectrum measured 
      in a rubidium vapor cell.
    }
    \label{fig2}

    \vspace{1em}

    \includegraphics[width=\linewidth,height=0.36\textheight,keepaspectratio]{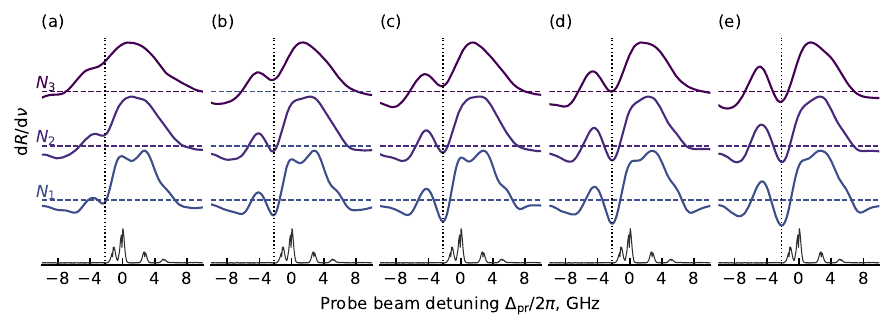}
    \caption{Derivatives $\mathrm{d}R/\mathrm{d}\nu$ of the selective reflection spectra 
      measured at a fixed pump-laser detuning of $-2.2$~GHz (indicated by the vertical 
      dotted lines) for different pump-beam intensities of (a)~2.6~kW\,cm$^{-2}$, 
      (b)~4.2~kW\,cm$^{-2}$, (c)~5.4~kW\,cm$^{-2}$, (d)~6.9~kW\,cm$^{-2}$, and 
      (e)~8.8~kW\,cm$^{-2}$.
      Each row corresponds to a different number density $N_i$, indicated on the left, 
      increasing from the lowest density, $N_1=1.2\times10^{17}$~cm$^{-3}$, 
      to the highest density, $N_3=2.5\times10^{17}$~cm$^{-3}$. 
      The bottom row shows the reference saturated absorption spectrum measured 
      in a rubidium vapor cell.
    }
    \label{fig3}
\end{figure*}

A spectroscopic study of selective reflection is performed with two tunable external cavity 
diode lasers, a weak probe laser and a powerful pump laser. 
The setup is the same as in our previous work~\cite{VASJQSRT2026}. 
The probe laser is scanned over the reflection spectra.
For calibration of the probe laser frequency, we record the saturated 
absorption spectrum of a reference vapor cell at room temperature. 
The frequencies of the lasers are measured with the wavemeter~\cite{SaakyanQE2015}. 
The detuning control allows us to keep the pump frequency within 20~MHz. 
The detuning of the pump laser is measured from the $5S_{1/2}(F = 3) - 5P_{3/2}(F' = 4)$
hyperfine component in $^{85}$Rb atoms.

The pump and probe optical beams with orthogonal linear polarizations are combined 
on a beam splitter cube and then focused on the inner surface of the input YAG window 
of the cell. 
The orthogonal polarizations of the beams are needed to eliminate the four-wave mixing 
at the window--vapor interface~\cite{VASFWMPRA1997}. 
The pump beam intensity $I$ is varied from zero to $8.8$~kW\,cm$^{-2}$.
The incidence angle of the probe beam is 64~mrad, and the incidence angle of the pump beam 
is 4~mrad. 
The reflected probe beam is sent to the photodetector. 
The signal from the photodetector is stored and analyzed by a digital oscilloscope 
and a computer. 
The normalized ratio $\delta R=(R-R_0)/R_0$ and the derivative 
$\mathrm{d}R/\mathrm{d}\nu$ are recorded. 
The reflection coefficient of the interface window--vacuum is $R_0=|(n-1)/(n+1)|^2 = 0.085$.
The spectral resolution is enhanced by using the frequency derivative of the 
reflection coefficient~\cite{VASJQSRT2026}. 
The selective reflection coefficient 
\begin{equation}\label{eq2}
  R= \Big|\frac{(n-\varepsilon)}{(n+\varepsilon)}\Big|^2
\end{equation}
can be calculated using the dielectric function $\varepsilon$ in the linear 
regime~\cite{BoydPRL1991}
\begin{equation}\label{eq3}
  \varepsilon=1+\sum_j\frac{k_j N}{(\omega-\omega_j)-\mathrm{i} \Gamma_0/2},
\end{equation}
where $k_j$ and $\omega_j$ are the factor and the resonance frequency for the corresponding 
$j$th hyperfine component.
Possible frequency shifts are not considered in the present paper. 
For simple estimations, we write the simplified relations 
for $\delta R$
\begin{equation}\label{eq4}
  \delta R \approx \mathrm{Re}(\varepsilon-1) \approx \sum_j\frac{k_jN(\omega-\omega_j)}
  {(\omega-\omega_j)^2+\Gamma_0^2/4},
\end{equation}
and for the frequency derivative $\mathrm{d}R/\mathrm{d}\nu$
\begin{equation}\label{eq5}
  \frac{dR}{d\nu}\approx\sum_jk_j\frac{\Gamma_0^2/4-(\omega-\omega_j)^2}
  {\big((\omega-\omega_j)^2+\Gamma_0^2/4\big)^2}
\end{equation}

\begin{figure}[t]
    \centering
    \includegraphics{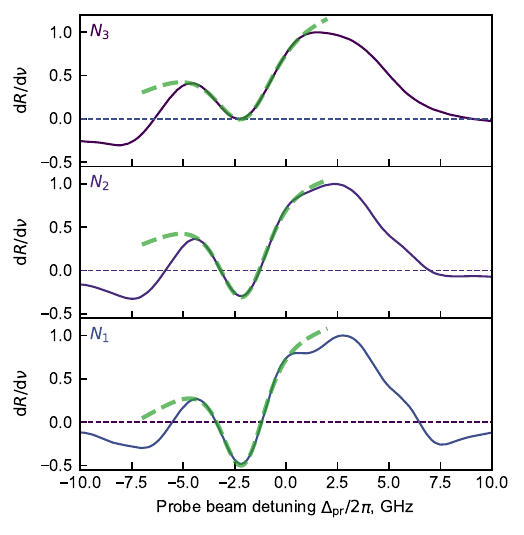}
    \caption{Representative fits of the saturation resonances for the derivative spectra
      measured at $I=6.9$~kW\,cm$^{-2}$ and a pump laser detuning of $-2.2$~GHz, 
      corresponding to panel (d) in Fig.~\ref{fig3}. 
      The three curves correspond to the atomic number densities $N_i$. 
      The dashed lines show the best fits obtained using Eq.~\ref{eq6}, with 
      $\Delta\nu_\mathrm{dip}$, $\gamma$, $A$, $p_0$, and $p_1$ treated as 
      free parameters. 
      The fitted parameters are $N_1$: $\{\Delta\nu_\mathrm{dip},\gamma,A,p_0,p_1\}
      =\{-2.10,2.14,-4.53,0.73,0.11\}$; $N_2$: $\{-2.10,2.42,-5.27,0.77,0.08\}$;
      and $N_3$: $\{-2.05,2.82,-5.48,0.89,0.10\}$.
      Here $\Delta\nu_\mathrm{dip}$ and $\gamma$ are given in GHz, $A$ in 
      arb.~u.~GHz$^2$, $p_0$ in arb.~u., and $p_1$ in arb.~u.\,GHz$^{-1}$.
    }
    \label{fig4}
\end{figure}

Selective reflection spectra, presented in Figures~\ref{fig1}-\ref{fig4}, are obtained under 
different experimental conditions. 
In each figure, "zero" of the probe laser detuning $\Delta_{\mathrm{pr}}/2\pi$
corresponds the $5S_{1/2}(F = 3)-5P_{3/2}(F'=4)$ hyperfine component of $^{85}$Rb
D$_2$-line.
In Fig.~\ref{fig1} the reflection spectra and reference absorption spectrum are shown.
According to Eq.~\eqref{eq5}, the characteristic width of the selective reflection spectra 
can be evaluated from the frequency interval $\Delta\nu_\mathrm{SR}$ between the two 
zero crossing points of the derivative $\mathrm{d}R/\mathrm{d}\nu$. 
In the experiment, these intervals are determined from the derivative spectra shown 
in Fig.~\ref{fig1}(b). 
For the three densities $N_1$, $N_2$, and $N_3$, the measured values are 
$\Delta\nu_\mathrm{SR}=17$, $22.5$, and $32$~GHz, respectively. 
These values exceed the self-broadened width $\Gamma_0/2\pi$ calculated with Eq.~\ref{eq1} 
by approximately $4$~GHz. 
This difference can be attributed to the influence of nearby hyperfine structure components 
on the recorded spectral profiles.

The nonlinear response of the selective reflection derivative spectra is shown in 
Figs.~\ref{fig2} and~\ref{fig3}. 
Figure~\ref{fig2} demonstrates the effect of pump-laser detuning on the saturation dips 
at the maximum pump intensity, $I=8.8$~kW\,cm$^{-2}$.
The curves in Fig.~\ref{fig2}(a) and (b) are obtained with off-resonance optical 
excitation, as in~\cite{VASPRL1996} and~\cite{VASreviewLPL2011}. 
Photon reabsorption and nonradiative excitation transfer induce incoherent 
saturation by the pump and spectral narrowing of all hyperfine components. 
Induced dips appear in the central parts of the reflection spectra. 
The curves in Fig.~\ref{fig2}(b, c, d) are obtained with the resonant optical excitation, 
as in~\cite{VASJQSRT2026}. 
The pump beam induces narrow saturation resonances. 
Only the isolated saturation resonances in Fig.~\ref{fig2}(b) coincides with the pump 
detuning $-2.2$~GHz.
The saturation resonances in Figs.~\ref{fig2}(c, d) are shifted and broadened due to 
the overlap of neighboring hyperfine components. 
For the subsequent fitting, we select the isolated contrast dip recorded at a pump-laser 
detuning of $-2.2$~GHz [Fig.~\ref{fig2}(b)]. 
This dip is used for the intensity dependent measurements shown in Fig.~\ref{fig3}. 
Fig.~\ref{fig3} shows the evolution of the saturation dip with increasing pump beam 
intensity at fixed pump laser detuning.
Figure~\ref{fig3} confirms the selection of the saturation resonances induced at 
pump laser detuning $-2.2$~GHz. 
Only at low intensities the dips are shifted and deformed. 
The shapes of such resonances are not discussed in the current paper. 
Here we focus on the resonances in the range of higher intensities.  

To determine the parameters of the saturation resonances, the isolated dips are fitted 
using the function
\begin{equation}\label{eq6}
  F_\mathrm{NL} = F_{\mathrm{slope}}(\Delta\nu)+F_{\mathrm{dip}}(\Delta\nu)
\end{equation}
which is proportional to the real part of a nonlinear dielectric function 
$\varepsilon_{\mathrm{NL}}$. 
Here, it is more convenient to use $\nu = \omega/2\pi$ instead of $\omega$, 
and $\Delta\nu=\Delta_\mathrm{pr}/2\pi$.
A function $F_\mathrm{slope}(\Delta\nu)$ is used to describe the slope of the derivative 
profile $\mathrm{d}R/\mathrm{d}\nu$, where the pump beam burned the dip. 
A function $F_\mathrm{slope}(\Delta\nu)$ is expressed as a polynomial function 
\begin{equation}\label{eq7}
  P_n(\Delta\nu)=p_0 + p_1 \Delta\nu + p_2 \Delta\nu^2 + p_3 \Delta \nu^3 + \cdots.  
\end{equation}
The dip can be fitted by a function
\begin{equation}\label{eq8}
  F_\mathrm{dip}(\Delta\nu)=-A\frac{(\gamma^2-(\Delta\nu-\Delta\nu_\mathrm{dip})^2)}
  {\big((\Delta\nu-\Delta\nu_\mathrm{dip})^2+\gamma^2\big)^2}.
\end{equation}
The fitting process is demonstrated in Fig.~\ref{fig4}.
The fitting procedure uses Eq.~\ref{eq6} with a first-order polynomial slope 
function, $F_\mathrm{slope}(\Delta\nu)=P_1(\Delta\nu)$. 
The coefficients $p_0$ and $p_1$, amplitude $A$, dip center frequency 
$\Delta\nu_\mathrm{dip}$, and dip width $\gamma$ are treated as free parameters.

\begin{figure}[t]
    \centering
    \includegraphics{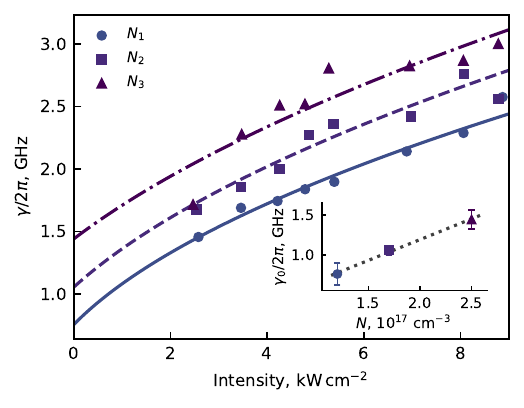}
    \caption{
      Power broadening of the saturation resonance width. 
      The dip width $\gamma$ extracted from fits using Eq.~\eqref{eq6} is plotted
      as a function of pump beam intensity $I$ for three atomic number densities $N_i$. 
      The solid lines show fits using Eq.~\eqref{eq10}, with $\gamma_0$ and 
      $I_\mathrm{sat}$ treated as free parameters. 
      The corresponding values of $\gamma_\mathrm{col}=2\gamma_0$ and $I_\mathrm{sat}$ 
      are listed in Table~\ref{tab}. 
      The inset shows the zero-intensity width $\gamma_0$ as a function of atomic number 
      density $N$, which increases linearly.
    }
    \label{fig5}
\end{figure}

The fitting function for power broadening of a homogeneously broadened transition according to
the textbook~\cite{Demtroderbook2008} can be written as 
\begin{equation}\label{eq9}
  \gamma = \gamma_0 \sqrt{1+\frac{E^2\mu^2}{\hbar^2\gamma_0^2}}.
\end{equation}
Here, $E$ is the optical electric field and $\mu$ is the atomic dipole moment. 
The intensity is given by $I = cE^2/8\pi$. 
Therefore, using the pump intensity $I$, the fitting function can be written as
\begin{equation}\label{eq10}
    \gamma=\gamma_0\sqrt{1+I/I_\mathrm{sat}},
\end{equation}
where $\gamma_0$ and $I_\mathrm{sat}$ are fitting parameters.
The power broadening of the fitted dip width $\gamma$ is shown in Fig.~\ref{fig5}. 
The data are fitted using Eq.~\eqref{eq10}, with $\gamma_0$ and $I_\mathrm{sat}$ 
as free parameters. 
The inset shows that the zero-intensity width $\gamma_0$ increases linearly with 
the atomic number density $N$. 
The linear dependence of the collision half-width in the inset is consistent with theoretical 
results in~\cite{MukamelPRA1994}. 

The results of the fit are shown in Table~\ref{tab}. 
We introduce a collision width $\gamma_\mathrm{col}=2\gamma_0$ (FWHM) for a better 
comparison with the line self-broadening. 
The ratio $\rho=\Gamma_0/\gamma_\mathrm{col}$ is presented in Table~\ref{tab}.

\begin{table}[ht]
  \centering
  \begin{threeparttable}
    \caption{Fitted and estimated parameters for different atomic densities.}
    \label{tab}

    \setlength{\tabcolsep}{3.2pt}

    \begin{tabular}{
      @{}
      c
      c
      c
      c
      c
      c
      @{}
    }
      \toprule
      {$i$} &
      \makecell{$N_i$,\\$10^{17}$ cm$^{-3}$} &
      \makecell{$\Gamma_0/2\pi$,\\ GHz \\ (FWHM)} &
      \makecell{$\gamma_\mathrm{col}/2\pi$, \\ GHz \\ (FWHM)} &
      \makecell{$I_\mathrm{sat}$\\kW\,cm$^{-2}$} &
      {$\rho$} \\
      \midrule
      1 & 1.2 & 13.2 & $1.52 \pm 0.28$ & $0.96 \pm 0.41$ & $8.68 \pm 2.43$ \\
      2 & 1.7 & 18.7 & $2.12 \pm 0.12$ & $1.05 \pm 0.22$ & $8.82 \pm 1.06$ \\
      3 & 2.5 & 27.5 & $2.88 \pm 0.24$ & $1.44 \pm 0.57$ & $9.55 \pm 2.29$ \\
      \bottomrule
    \end{tabular}
  \end{threeparttable}
\end{table}

Considering the error bars, the ratio $\rho$ can be treated as a fixed value 
regardless of the density $N_i$. 
In~\cite{MukamelPRA1994}, it is shown that the 
theoretical self-broadening $\Gamma$ is the sum of the static width 
$\Gamma_\mathrm{S}$ and the collision width $\Gamma_\mathrm{C}$ 
\begin{equation}\label{eq11}
  \Gamma=\Gamma_\mathrm{S}+\Gamma_\mathrm{C}=\Big(\frac{2}{3}\pi+\frac{\pi}{4}\Big)E_0=\frac{11}{12}\pi E_0,
\end{equation}
where $E_0 = 4\pi\mu^2N/3$. 
The theoretical ratio of the line self broadening $\Gamma$ to the collision
width $\Gamma_\mathrm{C}$ is 11:3 and is independent of density, which is 
consistent with our experimental results.

\section{Conclusions}
In the present work, we investigate the nonlinear spectra of selective reflection 
from high-density rubidium vapor using a pump-probe technique in the density range 
$(1.2-2.5)\times10^{17}$~cm$^{-3}$. 
Narrow saturation resonances are observed inside the self-broadened line shapes. 
By analyzing the shapes of saturation resonances, we estimate the collision width. 
The inhomogeneous line self-broadening is a combination of static and collision widths. 
The obtained experimental results can be explained by the theoretical model developed 
in~\cite{MukamelPRA1994}. 
In our previous publication~\cite{VASJQSRT2026}, we reported a transformation of the 
line self-broadening from inhomogeneous to homogeneous due to a density change from  
$2.5\times10^{17}$~cm$^{-3}$ to $3.6\times 10^{17}$~cm$^{-3}$. 
We explained the homogeneous broadening by quasi-molecular dipole-dipole 
interactions~\cite{VASZeemanPRA1997}. 
Additional experimental studies of the homogeneous broadened resonance transitions 
may help to develop an extended model of dipole-dipole interactions in a dense 
atomic gas. 
Absorption spectroscopy of dense atomic gases can be performed with ultrathin 
gas cells~\cite{SarkisyanPRL2012,SarkisyanPRL2018,sargsyan2025doppler}.

\section*{CRediT authorship contribution statement}
\textbf{Vladimir Sautenkov}: Conceptualization, Methodology, Validation, 
                            Writing -- original draft.
\textbf{Sergey Saakyan}: Investigation, Software, Validation, Formal analysis,      
                        Writing -- review \& editing. 
\textbf{Andrei Bobrov}: Formal analysis, Writing -- review \& editing. 
\textbf{Eugenia Vilshanskaya}: Formal analysis, Writing -- review \& editing. 
\textbf{Boris B. Zelener}: Supervision, Resources, Writing -- review \& editing,
                        Funding acquisition.

\section*{Declaration of Competing Interest}
The authors declare that they have no known competing financial interests or 
personal relationships that could have appeared to influence the work reported 
in this paper.

\section*{Data availability}
Data will be made available upon request.

\section*{Acknowledgements}
The research has been supported by the Ministry of Science and Higher Education 
of the Russian Federation (State Assignment No.\,075-00270-26-00).

\bibliographystyle{elsarticle-num}
\bibliography{refs}

@article{Lewis1980,
  title = {Collisional relaxation of atomic excited states, line broadening and
           interatomic interactions},
  author = {Lewis, E L},
  journal = {Phys Rep},
  volume = {58},
  number = {1},
  pages = {1--71},
  year = {1980},
  publisher = {Elsevier},
  doi = {10.1016/0370-1573(80)90056-3},
}

@article{BoydPRL1991,
  title = {Linear and nonlinear optical measurements of the \rm{L}orentz local
           field},
  author = {Maki, Jeffery J and Malcuit, Michelle S and Sipe, J E and Boyd,
            Robert W},
  journal = {Phys Rev Lett},
  volume = {67},
  number = {8},
  pages = {972},
  year = {1991},
  doi = {10.1103/PhysRevLett.67.972},
}

@article{MukamelPRA1994,
  title = {Self-broadening and exciton line shifts in gases: \rm{B}eyond the
           local-field approximation},
  author = {Leegwater, Jan A and Mukamel, Shaul},
  journal = {Phys Rev A},
  volume = {49},
  number = {1},
  pages = {146},
  year = {1994},
  doi = {10.1103/PhysRevA.49.146},
}

@article{MukamelJCP1991,
  title = {Photon echoes of polyatomic molecules in condensed phases},
  author = {Yan, Yi Jing and Mukamel, Shaul},
  journal = {J Chem Phys},
  volume = {94},
  number = {1},
  pages = {179--190},
  year = {1991},
  publisher = {American Institute of Physics},
  doi = {10.1063/1.460376},
}

@article{VASPRL1996,
  title = {Dipole-dipole broadened line shape in a partially excited dense
           atomic gas},
  author = {Sautenkov, V A and \rm{van Kampen}, H and Eliel, E R and Woerdman, J
            P},
  journal = {Phys Rev Lett},
  volume = {77},
  number = {16},
  pages = {3327},
  year = {1996},
  publisher = {APS},
  doi = {10.1103/PhysRevLett.77.3327},
}

@article{HolsteinPRA1951,
  title = {Imprisonment of resonance radiation in gases. II},
  author = {Holstein, T},
  journal = {Phys Rev},
  volume = {83},
  number = {6},
  pages = {1159},
  year = {1951},
  publisher = {APS},
  doi = {10.1103/PhysRev.83.1159},
}

@article{Biberman1947,
  title = {On the Theory of the Diffusion of Resonance Radiation; K teorii
           diffusii resonansnogo izluchenia},
  author = {Biberman, L M},
  journal = {Zhur. Eksptl. i Teoret. Fiz.},
  volume = {17},
  year = {1947},
  url = {http://refhub.elsevier.com/S0022-4073(20)30710-X/sbref0001},
}

@article{VASJQSRT2020,
  title = {Spectral dependence of nonlinear radiation trapping in high density
           atomic vapor},
  author = {Sautenkov, Vladimir and Saakyan, Sergey and Zelener, Boris B},
  journal = {J Quant Spectrosc Radiat Transf},
  volume = {256},
  pages = {107349},
  year = {2020},
  publisher = {Elsevier},
  doi = {10.1016/j.jqsrt.2020.107349},
}

@article{VASNSRJQSRT2024,
  title = {Spectroscopy of resonantly saturated selective reflection from
           high-density rubidium vapor using the pump-probe technique},
  author = {Sautenkov, Vladimir and Saakyan, Sergey and Bobrov, Andrei and
            Khalutornykh, Leonid and Zelener, Boris B},
  journal = {J Quant Spectrosc Radiat Transf},
  volume = {328},
  pages = {109153},
  year = {2024},
  publisher = {Elsevier},
  doi = {10.1016/j.jqsrt.2024.109153},
}

@book{Demtroderbook2008,
  title = {Laser spectroscopy: vol. 1 basic principles},
  author = {Demtr{\"o}der, Wolfgang},
  year = {2008},
  publisher = {Springer Berlin, Heidelberg},
  doi = {10.1007/978-3-540-73418-5},
}

@article{AdamsJPhysB2011,
  title = {Absolute absorption on the rubidium \rm{D}1 line including resonant
           dipole--dipole interactions},
  author = {Weller, Lee and Bettles, Robert J and Siddons, Paul and Adams,
            Charles S and Hughes, Ifan G},
  journal = {J Phys B},
  volume = {44},
  number = {19},
  pages = {195006},
  year = {2011},
  publisher = {IOP Publishing},
  doi = {10.1088/0953-4075/44/19/195006},
}

@article{VASFWMPRA1997,
  title = {Observation of narrow resonances inside homogeneously self-broadened
           lines in pump-probe reflection experiments},
  author = {Sautenkov, Vladimir A and Gamidov, RG and Weis, Antoine},
  journal = {Phys Rev A},
  volume = {55},
  number = {4},
  pages = {3137},
  year = {1997},
  publisher = {APS},
  doi = {10.1103/PhysRevA.55.3137},
}

@article{VASZeemanPRA1997,
  title = {Observation of collisional modification of the \rm{Z}eeman effect in
           a high-density atomic vapor},
  author = {\rm{van Kampen}, H and Papoyan, A V and Sautenkov, V A and
            Castermans, P H A M and Eliel, E R and Woerdman, J P},
  journal = {Phys Rev A},
  volume = {56},
  number = {1},
  pages = {310},
  year = {1997},
  publisher = {APS},
  doi = {10.1103/PhysRevA.56.310},
}

@article{SarkisyanPRL2012,
  title = {Cooperative \rm{L}amb shift in an atomic vapor layer of nanometer
           thickness},
  author = {Keaveney, James and Sargsyan, Armen and Krohn, Ulrich and Hughes,
            Ifan G and Sarkisyan, David and Adams, Charles S},
  journal = {Phys Rev Lett},
  volume = {108},
  number = {17},
  pages = {173601},
  year = {2012},
  publisher = {APS},
  doi = {10.1103/PhysRevLett.108.173601},
}

@article{SarkisyanPRL2018,
  title = {Collective \rm{L}amb shift of a nanoscale atomic vapor layer within a
           sapphire cavity},
  author = {Peyrot, Tom and Sortais, Yvan R P and Browaeys, A and Sargsyan,
            Armen and Sarkisyan, David and Keaveney, J and Hughes, I G and Adams,
            Charles S},
  journal = {Phys Rev Lett},
  volume = {120},
  number = {24},
  pages = {243401},
  year = {2018},
  publisher = {APS},
  doi = {10.1103/PhysRevLett.120.243401},
}

@article{VASreviewLPL2011,
  title = {Line shapes of atomic transitions in excited dense gas},
  author = {Sautenkov, V A},
  journal = {Laser Phys Lett},
  volume = {8},
  number = {11},
  pages = {771--781},
  year = {2011},
  doi = {10.1002/lapl.201110070},
}

@article{VASJETPLett2025,
  title = {Splitting of the Selective Reflection Spectrum Due to the Resonance
           Optical Saturation of a Dipole-Broadened Atomic Transition},
  author = {Sautenkov, V A and Saakyan, S A and Bobrov, A A and Zelener, B B},
  journal = {JETP Lett},
  volume = {122},
  number = {6},
  pages = {342--346},
  year = {2025},
  publisher = {Springer},
  doi = {10.1134/S0021364025607961},
}

@article{Cohen-TannoudjiJPhysB1977,
  title = {Dressed-atom description of resonance fluorescence and absorption
           spectra of a multi-level atom in an intense laser beam},
  author = {Cohen-Tannoudji, Claude and Reynaud, Serge},
  journal = {J. Phys. B: At. Mol. Phys.},
  volume = {10},
  number = {3},
  pages = {345--363},
  year = {1977},
  doi = {10.1088/0022-3700/10/3/005},
}

@article{VASJQSRT2026,
  title = {Optical-field-induced dips and splits in nonlinear spectra of
           selective reflection from high-density atomic vapor},
  author = {Sautenkov, Vladimir and Saakyan, Sergey and Bobrov, Andrei and
            Zelener, Boris B},
  journal = {J. Quant. Spectrosc. Radiat. Transf.},
  pages = {109796},
  year = {2025},
  publisher = {Elsevier},
  doi = {10.1016/j.jqsrt.2025.109796},
}

@article{SaakyanQE2015,
  title = {Frequency control of tunable lasers using a
           frequency-calibrated-meter in an experiment on preparation of Rydberg
           atoms in a magneto-optical trap},
  author = {Saakyan, Sergei Aramovich and Sautenkov, Vladimir Alekseevich and
            Vilshanskaya, Evginia Vladimirovna and Vasiliev, VV and Zelener,
            Boris Borisovich and Zelener, Boris Vigdorovich},
  journal = {Quantum Electron.},
  volume = {45},
  number = {9},
  pages = {828--832},
  year = {2015},
  publisher = {Turpion Ltd and the Russian Academy of Sciences},
  doi = {10.1070/QE2015v045n09ABEH015708},
}

@article{sargsyan2025doppler,
  title = {Doppler-free selective reflection spectroscopy of the $6s^2\rm{S}_{
           1/2} -7p^2\rm{P}_{3/2}$ transition of \rm{C}aesium using an optical
           nanocell},
  author = {Sargsyan, Armen and Momier, Rodolphe and Sarkisyan, David},
  journal = {J Phys B},
  volume = {58},
  number = {19},
  pages = {195001},
  year = {2025},
  publisher = {IOP Publishing},
  doi = {10.1088/1361-6455/ae0a98},
}

\end{document}